\documentclass[aps,prc,twocolumn,amssymb,floatfix]{revtex4}
\usepackage{graphicx}
\def\lsim{\mathrel{\rlap{
\lower4pt\hbox{\hskip-3pt$\sim$}}
    \raise1pt\hbox{$<$}}}     
\def\gsim{\mathrel{\rlap{
\lower4pt\hbox{\hskip-3pt$\sim$}}
    \raise1pt\hbox{$>$}}}     

\begin{document}
\title{Semi-central In+In collisions and Brown-Rho scaling}
\author{V.V.~Skokov}
\email{vvskokov@theor.jinr.ru}
\affiliation{Bogoliubov Laboratory of Theoretical Physics,
Joint Institute for Nuclear Research, \\
141980, Dubna, Russia}
\author{V.D.~Toneev}
\email{toneev@theor.jinr.ru}
\affiliation{Bogoliubov Laboratory of Theoretical Physics,
Joint Institute for Nuclear Research, \\
141980, Dubna, Russia}


\begin{abstract}
In connection with the claim made at the Quark Matter 2005
Conference that the Brown-Rho scaling is ruled out by NA60 data we
consider dimuon production from semi-central In+In collisions in a
full dynamical model. It is shown that if  only a modification of
the density-dependent $\rho$-mass is allowed, the maximum of
dimuon invariant mass spectra is only slightly below experimental
one. The additional inclusion of the temperature-dependent
modification factor  shifts the spectrum maximum toward lower
invariant masses making calculation results incompatible with
data. A further study is needed to disentangle the BR dropping
$\rho$ mass scaling and strong broadening.
\end{abstract}

\maketitle

Because of weak interaction, dileptons play an exceptional role
among different probes providing information on a state of highly
 compressed and hot nuclear matter formed in relativistic heavy ion
collisions. Generally, a dilepton yield depends on both global
properties of matter constituents (hadrons and/or quarks, gluons)
defined by the equation of state and  individual constituent
properties related to their in-medium modification. The analysis
of the $e^+e^-$ invariant mass spectra from central Pb+Au
collisions at the bombarding energy $E_{lab}$=158$A$ GeV, measured
by the CERES Collaboration, for certain  shows an excess radiation
in the range of invariant dilepton masses $0.2 \lsim M \lsim 0.7$
GeV. A possible interpretation of this excess was given in terms
of a strong in-medium $\rho$-meson modification (see review
articles~\cite{CB99,RW00}). Various scenarios of hadron
modification were proposed. However, low statistics, insufficient
mass resolution,  and large signal/background ratio in the CERES
experiments did not allow one to discriminate these scenarios, in
particular those based on the Brown-Rho (BR) scaling
hypothesis~\cite{BR91} assuming a dropping $\rho$ mass and on a
strong broadening as found in the many-body approach by Rapp and
Wambach~\cite{RW00,RW1}.

The proposed NA60 experiment with the muon detector, zero degree
calorimeter and the refined target area does not seem to suffer
from these CERES shortcomings. The first NA60 results for
$\mu^+\mu^-$ pair production in In+In collisions have recently
been presented at the Quark Matter 2005
Conference~\cite{NA60-QM05} (see also Fig.1 in Ref.~\cite{BR1})).
From comparison of these results with Rapp's theoretical
predictions it was declared that the ``BR scaling is ruled out by
NA60'' dimuon data whereas the many-body approach gives a
reasonable agreement. This claim immediately raised an objection
by Brown and Rho. In their two letters~\cite{BR1,BR2} they noted
three points that were neglected or simplified in the calculation
of the dropping mass scenario which make it having nothing to do
with that they believe BR scaling~\cite{BR91,BR04} to be.

All three key points stated in~\cite{BR1,BR2} may be essential in
the description  of dilepton production but still have not been
properly taken into account in the works available now in the
literature. This also concerns this letter~: We  assume the
validity of the vector dominance and do not discuss "sobar"
excitations. We basically concentrate on the influence of the
temperature and baryon density dependence in the form of scaling.
Moreover, nuclear interaction dynamics is presented in detail to
make clear which states mainly contribute to the observed dimuon
yield as well as the model parameters used.

The dynamics of heavy ion collisions is treated in terms of a
hybrid model where the initial interaction stage is described by
the transport Quark Gluon String Model (QGSM)~\cite{QGSM} and the
subsequent stage is considered as an  isoentropic expansion. The
latter stage is calculated within the relativistic 3D
hydrodynamics~\cite{ST05}  allowing different equations of state.
In our work, the mixed phase Equation of State
(EoS)~\cite{TFNFR04} is applied. This thermodynamically consistent
EoS uses the modified Zimanyi mean-field interaction for hadrons
and also includes interaction between hadron and quark-gluon
phases, which results in a cross-over deconfinement phase
transition. In addition to Ref.~\cite{TFNFR04}, the hard thermal
loop term was self-consistently added to the interaction of quarks
and gluons to get the correct asymptotics at $T>>T_c$ and
reasonable agreement of the model results with lattice QCD
calculations at finite temperature $T$ and chemical potential
$\mu_B$~\cite{latEoS}.

\begin{figure}[h]
  \includegraphics[width=7cm]{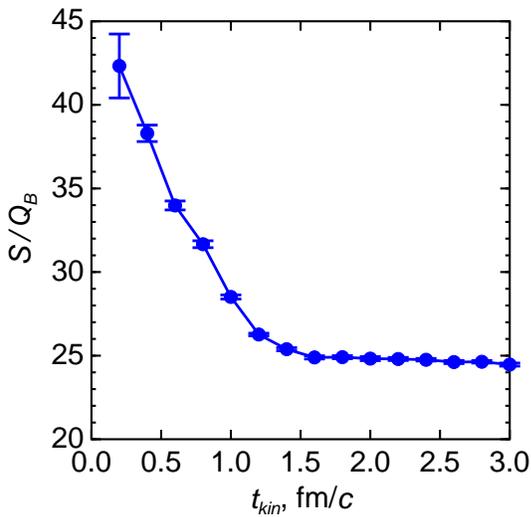}
  \caption{(Color online) Temporal dependence of entropy $S$ per baryon charge $Q_B$ of participants for
  semi-central In+In collision at $E_{lab}=$158$A$ GeV.}
\end{figure}

The ratio between  entropy $S$ and total participant baryon charge
$Q_B$, is shown in Fig.1 for In+In collisions at the impact
parameter $b=4$ fm and bombarding energy 158$A$ GeV estimated
within QGSM. Being calculated on a large 3D grid, this ratio is
less sensitive to particle fluctuation  as compared to entropy
itself. Small values of $Q_B$ at the very beginning of interaction
result in large values of the $S/Q_B$ ratio. It is clearly seen
that for $t_{kin}\gsim 1.3$ fm/$c$ this ratio is practically kept
constant and this stage may be considered as isoentropic
expansion.
\begin{figure}[t]
  \includegraphics[width=7cm]{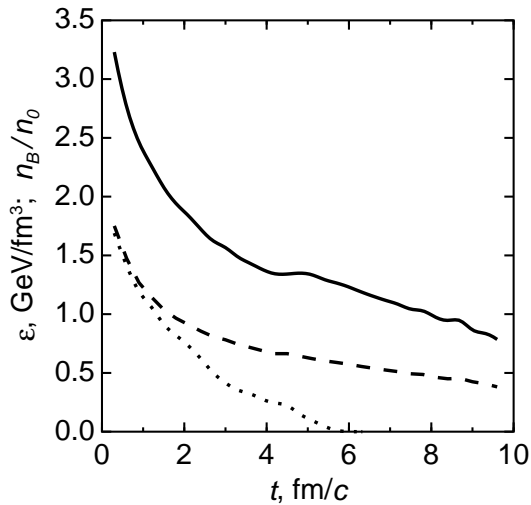}
  \caption{The average energy (solid line)  and baryon  (dashed) densities
  of an expanding  fireball formed in In+In collisions. Dotted line
  shows a contribution of quarks and gluons to the energy density.}
\end{figure}

 To proceed from kinetics to hydrodynamics, we  evaluate
conserving components of the energy-momentum tensor $T_{00},
T_{01}, T_{02}, T_{03}$ and baryon density (the zero component of
the baryon current) within QGSM at the moment $t_{kin}=1.3$ fm/$c$
in every cell on the 3D grid. This state is treated as an initial
state for subsequent hydrodynamic evolution of a fireball. The
time dependence of average thermodynamic quantities is presented
in Figs.2,3.

As  is seen, the average (over the whole excited system)  initial
energy density and the compression ratio, $n_B/n_0$, are not so
high, about 3.2 GeV/fm$^3$ and 1.7, respectively. These values are
by a factor of $\sim 2$ lower than those reached in central Pb+Au
collisions at the same energy in the CERES experiments. Both
quantities fall down in time and exhibit a weak structure related
to the softest point of the mixed phase EoS~\cite{TFNFR04}. The
depicted contribution of quark-qluon degrees of freedom to energy
density is not large and the muon creation from this phase will be
neglected below. The temperature evolution (Fig.3) looks quite
similarly. The calculated critical temperature at the considered
finite $\mu_B$ is about 160 MeV, so the evolution starts slightly
above this value. The end point of the temperature evolution curve
is about 135 MeV and is governed by freeze-out. In our model, the
freeze-out occurs locally and continuously during the  whole
evolution. The freeze-out condition is that the local energy
density (including 6 neighbor cells) is below a certain value
$\varepsilon_{fr}$ which was fixed by describing pion multiplicity
in central Pb+Pb collisions at the maximal SPS energy~\cite{ST05}.
One should note that the 1D Lorentz invariant Bjorken
hydrodynamics~\cite{B83} predicts an essentially faster fall down
and, therefore, a much shorter hydrodynamic evolution time.
\begin{figure}[b]
  \includegraphics[width=7cm]{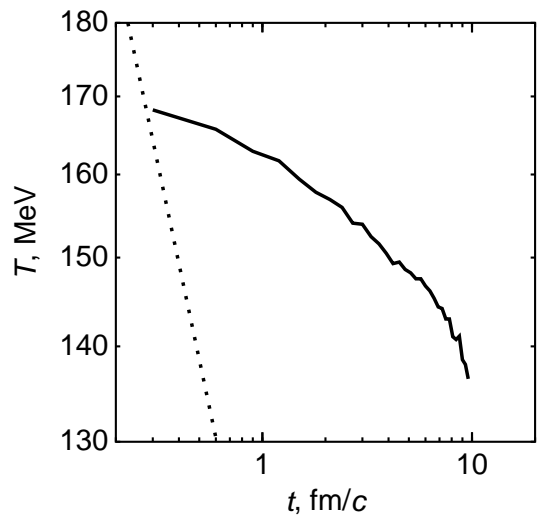}
  \caption{Evolution of the average temperature. The dotted line corresponds
  to the Bjorken regime with ultrarelativistic ideal gas EoS.}
\end{figure}

To find observable dilepton  characteristics, one should integrate
the emission rate over the whole time-space $x\equiv (t,\bf x)$
evolution, add the contribution from the freeze-out surface
(hadron cocktail), and take into account the experimental
acceptance. To simplify our task, we consider only the main
channel $\pi \pi \to l^+ l^-$. In this case the dilepton emission
rate is
\begin{eqnarray}
\frac{d^4 N}{d^4q} &=&-\int d^4 x \ {\cal L}(M) \
\frac{\alpha^2}{\pi^3 q^2} \ f_B(q_0,T(x)) \nonumber \\   &\times
{\rm Im}& \Pi_{em}(q,T(x),\mu_b(x)), \label{rate1}
\end{eqnarray}
where  the integration is carried out over the whole grid and time
from $t=0$ till the local freeze-out moment. Here
$q^2=M^2=q_0^2-{\mathbf q}^2$, the Bose distribution function is
defined as
\begin{equation}
 f_B(q_0, T(x)) = (e^{q_0/T(x)} - 1)^{-1}
\label{bose_func1}
\end{equation}
and lepton kinematic factor is
\begin{equation}
{\cal L}(M) = \left(1+2\frac{m_l^2}{M^2}\right)
\sqrt{1-4\frac{m_l^2}{M^2}} \label{massive_factor1}
\end{equation}
with the lepton mass $m_l$. If the  pole shift is
 neglected ($m_{\rho}^{pole} = m_\rho$), the imaginary part of the
electromagnetic current correlation function, integrated over $\bf
q$, for free  $\pi$$\pi$ annihilation  reads
\begin{equation}
{\rm Im} \Pi_{em}(M) = \frac{m^4_\rho}{g^2}  \frac{ {\rm Im} \Pi }
{ (M^2-m^2_\rho)^2 + ({\rm Im} \Pi)^2  }. \label{Rho_im1}
\end{equation}
The imaginary part of the $\rho$-meson self-energy in the one-loop
approximation  is
\begin{eqnarray}
 {\rm Im} \Pi = - \frac{g^2_{\rho \pi \pi}} {48 \pi} \frac{( M^2 - 4
m_\pi^2 )^{3/2}}{M}~, \label{ImPi}
\end{eqnarray}
where the parameters are defined in vacuum~\cite{KKW96}:
\begin{equation}
g_{\rho \pi \pi}  = 6.05; \,\, g = 5.03, \,\, m_{\rho} = 770 \
\mbox{MeV}. \label{params}
\end{equation}

To simulate the Brown-Rho scaling, we modify the in-medium masses
in eq.(\ref{Rho_im1}),(\ref{ImPi}) estimated according to the QCD
sum rules by Hatsuda and Lee~\cite{Hatsuda,hm}
\begin{equation}
m_{\rho}  \to m^*_{\rho}(x) =  m_{\rho} (1-0.15 \cdot n_B(x)/n_0)
\label{droppingmasses}
\end{equation}
and simultaneously apply the same modification to the vector
dominance coupling $m^2_\rho/g \to m^{*2}_\rho/g^*$~\cite{hm1}, as
it was suggested in~\cite{BR1,BR2}.

Rapp's results cited in ref.\cite{NA60-QM05} were obtained using a
different assumption on the in-medium mass~\cite{RW00}
\begin{equation}
m^*_{\rho}(x) =  m_{\rho} \left( 1-0.15 \cdot
\frac{n_B(x)}{n_0}\right)
\left(1-\left[\frac{T(x)}{T_c}\right]^2\right)^{0.3}~.
\label{Tdep_mass}
\end{equation}
The temperature dependence in eq.(\ref{Tdep_mass}) is motivated by
the T-dependence of quark condensate.

Dimuon invariant mass spectra from In+In collisions are presented
in Fig.4  for the $\pi\pi$ channel. Only the muon rapidity cut $3
< y_{lab} < 4$ is taken into account in our calculations. The NA60
experimental points demonstrate quite a high resolution in dimuon
mass; however, their absolute normalization is lost.  As is seen,
if the $T$-dependent dropping mass (\ref{Tdep_mass}) is assumed,
our curve is strongly shifted toward the low-mass region which is
in qualitative agreement with the Rapp result~\cite{NA60-QM05}. In
the case of the density dependent $\rho$ mass scaling
(\ref{droppingmasses}) the maximum position is only slightly below
experimental ones and this phenomenological scenario cannot
unambiguously be "ruled out". To compare properly dileptons from
the pion annihilation with experiment, the contribution of free
$\rho$-meson decay should be added. A number of these
$\rho$-mesons was estimated as thermodynamical emission from
frozen-out cells, according to the general procedure described
in~\cite{IRT05}. In the last case the agreement with experiment
will be even better if the muon contribution from the free $\rho$
decay is taken into account. Note that the calculated solid and
dotted lines in Fig.4 are multiplied by the same normalization
factor and, therefore, the shown ratio between the signal and free
$\rho$-meson decay is kept. Muon emission from pion annihilation
in the vacuum case very nearly follows the measured points;
however, as noted above, the observable quantity is not this
spectrum but its sum with the hadron decay cocktail (dotted line
in Fig.4). It is of interest that making use of the not modified
$m^2_\rho/g$ ratio increases the muon yield by a factor of $\sim$
2, as expected by~\cite{BR1,BR2}, but it is not visible in
normalized data because the spectrum shape is weakly changed.
\begin{figure}[t]
  \includegraphics[width=7cm]{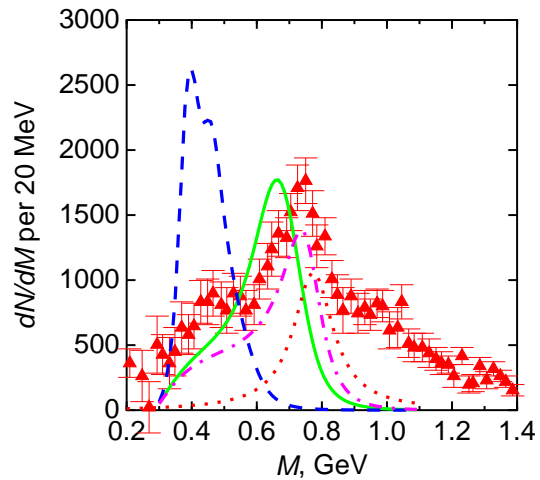}
  \caption{(Color online) Invariant mass distribution of dimuons from semi-central
  In+In collisions at the beam energy 158 $A$GeV. Experimental points are
  from~\cite{NA60-QM05}. Solid and dashed curves are calculated using the $\rho$-mass
  modification factors (\ref{droppingmasses}) and (\ref{Tdep_mass}), respectively.
  The dash-dotted curve neglects any in-medium modification.
  The dotted line indicates the hydrodynamically
  calculated $\rho$-meson decay at the freeze-out.}
\end{figure}

Finally, the in-medium modification of the $\rho$-meson mass based
on the $T$-dependent scaling~(\ref{Tdep_mass}) is hardly
compatible with the NA60 data. As noted in~\cite{BR1,BR2}, a more
realistic case corresponds to neglecting any temperature
dependence. This is indeed so, as follows from our results.  To
disentangle BR dropping $\rho$ mass scaling~\cite{BR91} and strong
broadening~\cite{RW00,RW1}, further detailed theoretical
investigations (e.g. see~\cite{RRM05}) and a comparison with
measured absolute muon yield  taking properly into account the
experimental acceptance are needed.

We are grateful to Yu.~B.~Ivanov and V.~N.~Russkih for numerous
discussions of details of hydrodynamical code, freeze-out
procedure, and account for experimental acceptance. We are
thankful to R.~Rapp for critical remarks  and S.~Damjanovic for
discussion of experimental data. One of us (VS) thanks organizers
of the QM2005 conference for making his participation possible.
This work was supported in part by DFG (project 436 RUS
113/558/0-3) and RFBR (grant 06-02-04001).


\begin{thebibliography}{99}
\bibitem{hm}
To be precise, we note that  Hatsuda and Lee result was
$m^*_{\rho} =  m_{\rho} (1-0.18 \cdot n_B/n_0)$. In our work  the
factor $0.15$ is used instead of  $0.18$ to be consistent
with~\cite{NA60-QM05}.
\bibitem{hm1}Note that this substituton really means some
suppression of the vector dominance~\cite{BR1,BR2,BR04}.
\end{thebibliography}
\end{document}